Article

# Ultrafast X-ray imaging of the light-induced phase transition in VO$_2$

Allan S. Johnson ®[1] ✉, Daniel Perez-Salinas[1], Khalid M. Siddiqui[2], Sungwon Kim[3], Sungwook Choi ®[3], Klara Volckaert[2], Paulina E. Majchrzak ®[2], Søren Ulstrup ®[2], Naman Agarwal[2], Kent Hallman[4], Richard F. Haglund Jr ®[4], Christian M. Günther ®[5], Bastian Pfau ®[6], Stefan Eisebitt[6,7], Dirk Backes ®[8], Francesco Maccherozzi[8], Ann Fitzpatrick[8], Sarnjeet S. Dhesi ®[8], Pierluigi Gargiani ®[9], Manuel Valvidares ®[9], Nongnuch Artrith ®[10], Frank de Groot[10], Hyeongi Choi[11], Dogeun Jang[11], Abhishek Katoch[11], Soonnam Kwon[11], Sang Han Park[11], Hyunjung Kim ®[3] & Simon E. Wall ®[2] ✉

Using light to control transient phases in quantum materials is an emerging route to engineer new properties and functionality, with both thermal and non-thermal phases observed out of equilibrium. Transient phases are expected to be heterogeneous, either through photo-generated domain growth or by generating topological defects, and this impacts the dynamics of the system. However, this nanoscale heterogeneity has not been directly observed. Here we use time- and spectrally resolved coherent X-ray imaging to track the prototypical light-induced insulator-to-metal phase transition in vanadium dioxide on the nanoscale with femtosecond time resolution. We show that the early-time dynamics are independent of the initial spatial heterogeneity and observe a 200 fs switch to the metallic phase. A heterogeneous response emerges only after hundreds of picoseconds. Through spectroscopic imaging, we reveal that the transient metallic phase is a highly orthorhombically strained rutile metallic phase, an interpretation that is in contrast to those based on spatially averaged probes. Our results demonstrate the critical importance of spatially and spectrally resolved measurements for understanding and interpreting the transient phases of quantum materials.

Direct observation of the nucleation and growth of a new phase is the solid-state equivalent of the molecular movie in chemistry[1]. However, this is challenging to realize in solids due to the range of length- and timescales involved, with dynamics occurring from the atomic scale up to the macroscale and from femtoseconds to microseconds.

Remarkable progress has been made in understanding atomic-scale dynamics on the ultrafast time scale by using diffraction-based probes, which can reveal atomic dynamics beyond the mean response[2] or dynamics of the average structure over nanoscale regions[3]. However, electronic probes are needed to understand the functionality of these states. This is crucial in quantum materials, where transient states can be created by light that have electronic properties not found in equilibrium[4–9]. In many cases, transient states are believed to be heterogeneous at the nanoscale[10–15], both because of the inhomogeneous excitation profiles generated by the pump beam in the depth of the material and due to the intrinsic heterogeneity of many quantum materials. A key challenge to understanding these phases is therefore to isolate the photo-induced state and to directly probe its properties at the nanoscale. Resonant coherent diffraction from electronically ordered states has been used to infer the statistical properties of a domain[16–18], but real-space images have not been produced. Here we use time- and energy-resolved coherent resonant soft X-ray imaging to observe the ultrafast insulator–metal phase transition in the prototypical quantum material vanadium dioxide ($VO_2$), over a macroscopic area with sub-50 nm spatial resolution and 150 fs time resolution, returning full spectroscopic information on transient states at the nanoscale.

[1]ICFO-Institut de Ciències Fotòniques, The Barcelona Institute of Science and Technology, Barcelona, Spain. [2]Department of Physics and Astronomy, Aarhus University, Aarhus, Denmark. [3]Department of Physics, Sogang University, Seoul, Korea. [4]Department of Physics and Astronomy, Vanderbilt University, Nashville, TN, USA. [5]Zentraleinrichtung Elektronenmikroskopie (ZELMI), Technische Universität Berlin, Berlin, Germany. [6]Max-Born-Institut, Berlin, Germany. [7]Institut für Optik und Atomare Physik, Technische Universität Berlin, Berlin, Germany. [8]Diamond Light Source, Didcot, UK. [9]ALBA Synchrotron Light Source, Barcelona, Spain. [10]Materials Chemistry and Catalysis, Utrecht University, Utrecht, Netherlands. [11]Pohang Accelerator Laboratory, Pohang, Korea. ✉e-mail: allan.s.johnson@gmail.com; simon.wall@phys.au.dk

The light-induced phase transition in $VO_2$ has been particularly influential in our understanding of optically driven quantum materials. At room temperature, $VO_2$ is in a monoclinic insulating (M1) phase characterized by dimerized pairs of vanadium ions. Light can break these dimers and drive the transition on the ultrafast timescale to the high-temperature rutile metallic (R) phase. The study of this transition has driven the development of multiple new techniques: it was the first solid–solid transition to be tracked by time-resolved X-ray diffraction[19], while ultrafast X-ray absorption techniques were first used to understand its electronic nature[20,21]. The transition to the rutile phase proceeds by disordering of the vanadium pairs on a sub-100 fs timescale[2], with the bandgap collapsing on a similar timescale[22]. However, it remains an open question whether the bandgap collapses before, or because of, the structural transition to the rutile phase.

In addition to these homogeneous effects, heterogeneity in the transient state is believed to play a key role in the dynamics[8–10,23–25]. Although the lattice and electronic properties change on the ultrafast timescale, analysis of the terahertz (THz) conductivity suggests that the rutile metallic phase locally nucleates and grows on a timescale of tens[10] to hundreds[25] of picoseconds. In addition, electron diffraction results suggest that an additional, meta-stable heterogeneous monoclinic metallic phase can form that can persist for hundreds of picoseconds[8,23] or even microseconds[9]. This phase is distinct from the transient state that occurs within the first 100 fs described above and is stabilized by a structural distortion that preserves the monoclinic symmetry. However, the existence of non-equilibrium phases remains debated[26,27] because a direct measurement of the metastable monoclinic metallic state has not been made.

To address the role of nanoscale heterogeneity and phase separation, we use time- and spectrally resolved resonant soft X-ray coherent imaging at the Pohang Accelerator Laboratory X-ray Free Electron Laser (PAL-XFEL)[28–31] to image the light-induced phase transition on the ultrafast timescale with nanometre spatial resolution. The power of this technique lies in the fact that it is a wide-field imaging technique that can exploit resonant X-ray spectroscopy both to provide a contrast mechanism between phases[32] and to enable the extraction of quantitative spectral information to aid phase identification on the nanoscale[33]. We report time-resolved imaging using two modes of operation, Fourier transform holography (FTH) and coherent diffractive imaging (CDI). In FTH, scattering patterns are inverted directly through the use of a fast Fourier transform[32] and require a single exposure. This enables rapid data collection but comes at the expense of losing the absolute values of the complex transmission. CDI, conversely, uses multiple exposures to increase the dynamic range of detected scattering patterns, with images obtained via iterative phase retrieval algorithms that yield the quantitative absolute transmission of the sample[33].

Figure 1a shows a multi-energy FTH image of $VO_2$ recorded at the vanadium L edge (517 eV, red channel) and oxygen K edge (529.5 eV and 531.25 eV, blue and green channel, respectively) at 325 K, a temperature at which the insulating and metallic phases coexist[32]. At these energies, the absorption coefficient shows large changes depending on the phase of the material, with the 529.5 eV signal showing a decrease in transmission and the 531.25 eV signal an increase when the system changes from M1 to R[34,35]. These changes can be used to provide contrast in imaging. The red–green–blue image of Fig. 1a shows the full topography of the sample, which comprises a range of crystallite sizes and grain boundaries (white). These boundaries are known to pin the position of the metallic domains[32], and in larger crystallites, clear domain coexistence of M1 (purple) and rutile metallic (green) phases can be seen. This pinning is vital as it ensures the initial domain structure recovers after photoexcitation (Supplementary Note 1 and Extended Data Figs. 1 and 2).

In Fig. 1b, we further verify the domain assignment by performing a temperature-dependent measurement. By subtracting the green and blue channels from the red–green–blue image, we can remove the temperature-independent topography and highlight the domain structure, as the transmission at 529.5 eV (blue channel) decreases in the metallic phase, while the transmission at 531.5 eV increases (green channel). Here we clearly see the R phase nucleating and forming a stripe state with M1, which is common in nanocrystals[32].

We now examine the dynamics of the heterogeneous state at 325 K, to observe the disappearance of the M1 phase, the growth of the R phase and/or the nucleation of new transient phases. We excite the system with 24 mJ cm$^{-2}$, 800 nm pulses and perform time-resolved FTH imaging at 529.5 eV photon energy. At this fluence, we are in the saturation regime, where increasing fluence no longer results in substantial changes, but the initial domain structure still recovers between photo-excitation events (Supplementary Note 1 and Extended Data Figs. 1 and 2). Large changes are found all over the sample, and in Fig. 1c, we focus on two regions of interest that are cuts across approximately 50-nm-wide rutile metallic domains surrounded by the M1 phase, which appear as dips in the image intensity. In both cases, the contrast between the metal and insulating states is strongly, but not completely, lost within the first 150 fs, after which any further substantial changes are only observed on the hundreds of picoseconds timescale.

We next examine the spatial dependence of the dynamics more closely. Figure 2a shows the pump-induced changes in the domain structure across the full field of view measured at 529.5 eV, relative to the state of the sample at −8.5 ps. Regions of increased transmission are shown in red, with decreased transmission in blue. Changes are observed across the entire sample with nanoscale texturing, but the spatial dependence of the pattern is roughly independent of pump–probe delay after the initial changes. A comparison with the static domain pattern (Fig. 2b) shows that the regions where the transmission decreased correspond to regions that began in the M1 phase, as expected for a transition to the R phase. However, regions that began in the R phase show an increase in transmission of similar magnitude to the changes seen in M1. This is unexpected because, although there may be excited state effects in R, these should be much weaker than the changes of M1 to R (ref. [26]).

Instead, the changes observed at the R phase are mainly artificial, resulting from the loss of the d.c. component in the FTH images, which can cause correlated dynamics across the whole image (Supplementary Note 2 and Extended Data Fig. 3). Therefore, to identify how many unique processes are occurring spatially, we perform a principal component analysis of the dynamic images. This process breaks down the transmission dynamics $T$ into a series of 'eigen' spatial, $A_i(x,y)$, and temporal, $f_i(t)$, functions of the form $T(x,y,t) = \sum_i A_i(x,y) f_i(t)$.

For times up to 20 ps, we find that only a single principal component is needed to describe the dynamical evolution of the images, meaning all regions in space evolve with the same temporal dynamic and the transmitted intensity can be represented as $T(x,y,t) = A(x,y)f(t)$. Only when data beyond approximately 100 ps are included in the analysis are additional terms needed (Extended

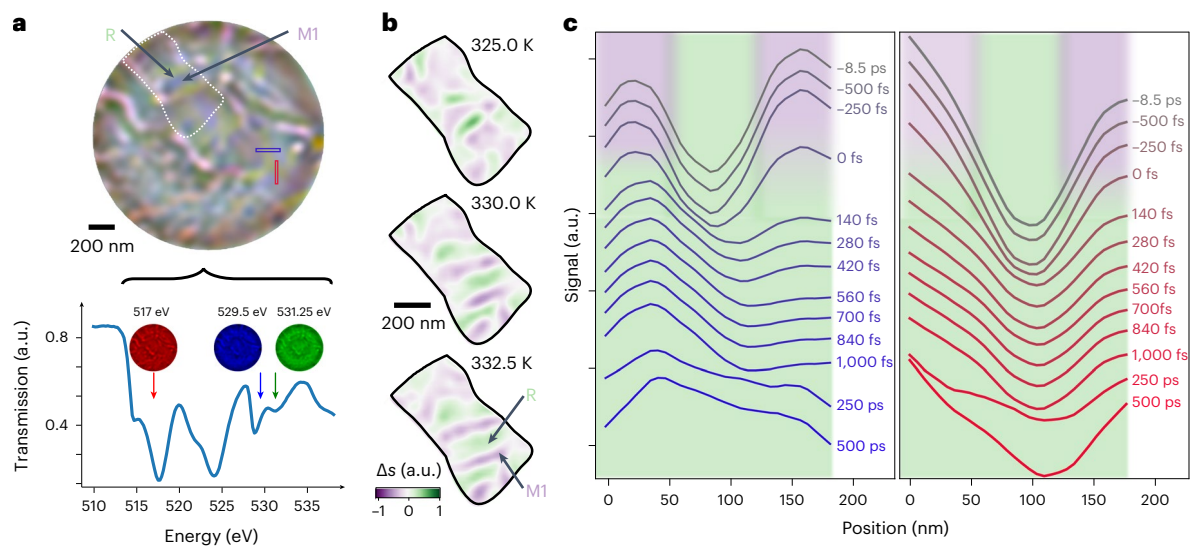

**Fig. 1 | Time-dependent X-ray holographic imaging of VO$_2$. a**, False colour composite FTH image of VO$_2$ from images recorded on the VO$_2$ soft X-ray resonance (bottom) at 517 eV (red), 529.5 eV (blue) and 531.25 eV (green). The metallic R phase appears green and insulating M1 phase appears purple. **b**, Temperature-dependent domain growth highlighted through the subtraction of the blue and green channels, Δs, which removes the sample morphology. The region of interest used is indicated by the white dotted region in **a**. **c**, Transmission dynamics of two line-outs spanning R regions surrounded by the M1 phase. Their positions are indicated in **a** and colour-coded. The domain structure, initially ~50 nm, is promptly lost. Background is shaded according to state of the material as a guide to the eye.

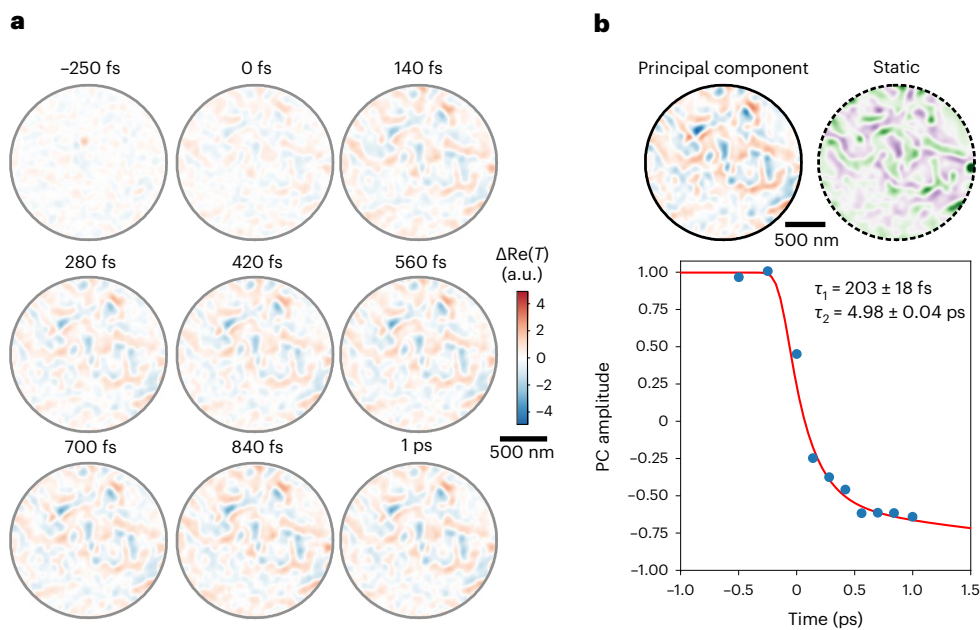

**Fig. 2 | Spatial dependence of the insulator-to-metal transition. a**, Full-field FTH images of pump-induced change in the real part (Re) of the transmission (T) images as a function of pump–probe delay. **b**, Comparison of the principal component (PC) obtained from **a** with a static two-colour difference image (529.5 eV and 531.25 eV) showing that regions that show increased (decreased) transmission correspond to regions starting in the green R (purple M1) phase. These early-time dynamics can be modelled as a double exponential decay with time constants $\tau_1 = 203 \pm 18$ fs and $\tau_2 = 4.98 \pm 0.04$ ps, corresponding to a near-resolution limited drop and a slower evolution respectively (see methods).

Data Figs. 4–6 and Supplementary Notes 3 and 4). The spatial and temporal response of the initial dynamics are plotted in Fig. 2b. The spatial pattern shows a correlation with the initial domain structure, demonstrating that the observed dynamics are indeed the result of the M1 regions of the sample switching to the metallic R phase and that no other local dynamics occur. A fit to the time trace for this process reveals two time constants, one, a near-resolution-limited $203 \pm 18$ fs fall time is consistent with the ultrafast nature of the structural[2] and electronic[21,22] changes during the M1 to R phase transition, while the second $4.98 \pm 0.04$ ps time constant is much slower. Critically, as only one principal component is needed to describe the dynamics, these two timescales occur in identical regions of the sample. This secondary picosecond time constant has been seen in multiple experiments, which have shown it to be fluence dependent[8,10,23,36,37]. However, the interpretation of this time constant has been debated. In some cases, it has been taken as evidence for a non-equilibrium, monoclinic metallic phase[8,23], while others have interpreted it as nucleation and growth of the metallic phase[10]. In the former case, the fast timescale is attributed to regions of the sample that undergo a direct transition from the M1 phase to the R phase, while the slower time constant is attributed to regions of the sample that transition from M1 to monoclinic insulating phases. This contrasts with our observations, which show both time constants occur in all regions of the sample. Similarly, in the nucleation and growth picture, the fast timescale should only be observed at the initial nucleation site; in addition, domain growth would require multiple principal components to describe. Therefore, both these scenarios are at odds with the data presented here, and thus we require an alternative description of the ultrafast phase transition.

To better understand the nature of the transient state formed after the picosecond evolution, we use spectrally resolved CDI to recover the full spectrum of the newly switched regions[33]. We acquire a hyperspectral image by scanning the probe wavelength, with images taken at 31 photon energies across the oxygen K edge at a delay of 20 ps after photoexcitation (Methods). The resulting spectrally integrated image is shown in Fig. 3a, which shows that the sample is remarkably homogeneous after excitation. However, as already noted in Fig. 1, markers of the initial phase are still observed at key energies. To elucidate the origin of this effect, we extract the transient spectra from all regions of the sample that were initially insulating and compare them with those that were initially metallic (Fig. 3b), enabling us to understand how the transient metallic state differs from the thermal state. The resulting spectra, and the differential, are shown in Fig. 3c,d.

Both regions are remarkably similar, and the resulting difference, less than 1% at 530.5 eV, is much smaller than the changes found from the insulator–metal transition, which are of order 10% (ref. [33]), showing the regions that were initially M1 have switched to the R phase. While the differences are smaller than the overall error on either spectra independently (Fig. 3c), the primary source of noise is fluctuations in the X-ray intensity, which are highly correlated across the sample and thus do not affect the differential spectra. The uncertainties on the difference spectra are reported independently in Fig. 3d and are of order 0.1% (Methods and ref. [33]). The spectral signatures observed are not consistent with thermal differences in the metallic phase[26], and, instead, we suggest that these differences result from strain generated during the phase transition. In equilibrium, the volume of $VO_2$ increases by approximately 1% between the M1 and R phase[38], but, on the ultrafast timescale, dimerization is lost without volume change, as the volume expansion can only occur after strain propagation. As a result, the photo-generated metallic regions are more strained than regions that were initially metallic.

A strain discontinuity at a surface leads to a strain wave[39,40]. The finite penetration depth of the pump causes the film to partially transform to the R phase, starting from the surface–vacuum interface. Fast moving, short-lived strain waves will be launched from the vacuum/sample and photo-generated R/M1 phase interfaces into the out-of-plane direction. As the excitation fluence is increased, the R phase will be switched deeper into the material, and the acoustic dynamics will speed up due to the higher speed of sound in the R phase[10]. When uniformly switched to the rutile phase at high fluence, strain waves will cross our sample in around 7 ps, consistent with our observed timescale (Supplementary Note 5 and Extended Data Fig. 7). The predicted speed increase with fluence is consistent with previous measurements[23]. In-plane strain relaxes more slowly because of the length scales involved (microns versus nanometres) and may be responsible for the spatially dependent dynamics observed at the hundreds of picosecond timescale.

The picture that emerges for the phase transition is shown in Fig. 3e. Diffraction measurements have shown that photoexcitation first breaks the vanadium dimers within approximately 100 fs[2], resulting in a strained rutile phase. Because of the anisotropy of the initial M1 phase, the degeneracy between the rutile *a* and *b* axis is broken, giving an orthorhombic structure. After approximately 10 ps, the out-of-plane strain relaxes, but the in-plane strain remains, preserving the orthorhombic nature and leading to the long-lived state shown in Fig. 3c,d. The tetragonal structure is only reached after hundreds of picoseconds when the in-plane strain can relax.

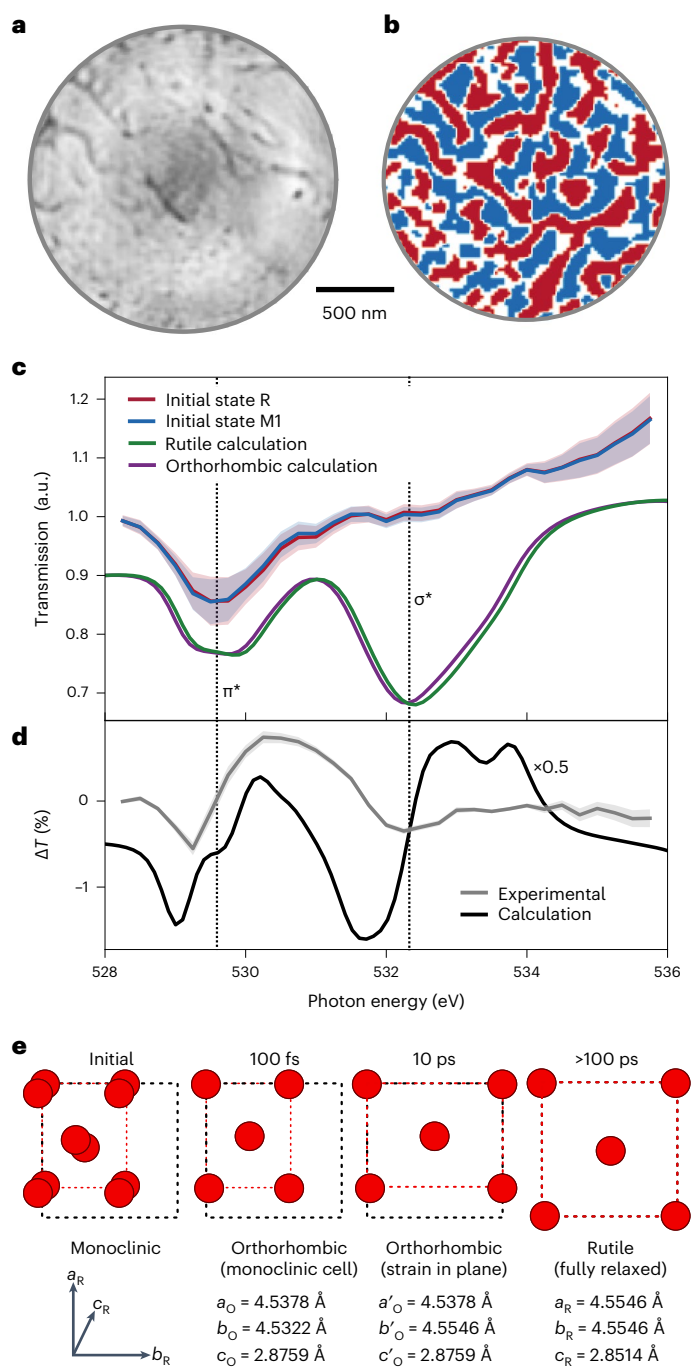

**Fig. 3 | Nanoscale X-ray spectroscopy of transient phases. a**, Frequency integrated hyperspectral CDI image at 20 ps pump–probe delay, showing clear topography but no signs of domains. **b**, Masks dividing the sample into regions that started in the R phase (red) and thus have not switched phase and in M1 phase (blue). **c**, Average transmission spectrum of blue and red regions at 20 ps delay. Shaded regions correspond to one standard deviation across all sample points. Also shown are calculated XAS spectra for a rutile metallic phase and an orthorhombically strained metallic phase (offset vertically). **d**, The measured difference in transmission (grey) between initially rutile and monoclinic regions (shaded regions correspond to one standard error) and the calculated value (black, offset vertically). **e**, Schematic evolution of the unit cell in $VO_2$ following photoexcitation (not to scale). The system relaxes the vanadium dimers within the first 100 fs[2], but cannot relax the strain on the same timescale. First the system is in a highly strained orthorhombic configuration, which preferentially relaxes along the out-of-plane direction (rutile $b$ axis) within 20 ps, before relaxing the in-plane strain to give the rutile structure.

To test this hypothesis, we have simulated the x-ray absorption spectra (XAS) of both the equilibrium rutile phase and the 20 ps orthorhombically strained configuration (Methods). The resulting spectra and differences are also plotted in Fig. 3 and show remarkable agreement with the experimental data. Qualitatively, the effect in both simulation and experiment is that the perturbed spectra are shifted in energy with respect to the initial state, but without changing their shape. Such a global shift, or 'shear', results in the difference spectra resembling the derivative of the spectra with respect to energy. The signs of the shifts are consistent with our interpretation, with the strained and switched spectra both moving to lower energies than the unstrained and unswitched spectra. There is also good agreement quantitatively at the $\pi^*$ state, where the two difference spectra match within a factor of 2. At the $\sigma^*$ state, the quantitative agreement is worse, although the qualitative agreement remains good. This results from the fact that all functionals used here overestimate the strength of the $\sigma^*$ state in the initial state compared with our data, which in turn influences the magnitude of the difference (Extended Data Fig. 8).

The implications of these results are twofold. First, we find no evidence for a heterogeneous monoclinic metallic phase for the fluence measured here. The strained orthorhombic state is qualitatively different from previously proposed out-of-equilibrium monoclinic phases[8,9,23,41], which result from decoupled internal degrees of freedom and neglect strain, but strain effects may explain the long-lived diffraction features previously associated with them[8,9,23]. In addition, the lack of nucleation and growth of the rutile metallic phase suggests that previously observed dynamics of the THz conductivity could result from the fact that strained $VO_2$ has a higher resistivity than the fully relaxed state[10,25].

We note that we cannot conclusively rule out the existence of the monoclinic metallic state in a different parameter range, as the proposed volume fraction generated is fluence dependent. While we are within the fluence regime in which a monoclinic metallic state has previously been claimed to exist[8,9,23], our fluence is higher than the fluence that is claimed to produce the maximum volume fraction of monoclinic metal[23]. Furthermore, our own measurements have shown that the fluence transition threshold in $VO_2$ is strongly sample and geometry dependent[26], and thus systematic fluence-dependent studies will be essential to definitively settle this question. In addition, although nucleation and growth dynamics have not been observed here, they may occur closer to the critical fluence, which is at much lower fluences than used here. As thin film samples show a spatially dependent critical temperature ($T_c$)[32,33] and the threshold fluence is known to correlate with $T_c$ (ref. [26]), one would expect that close to threshold excitation will only switch nanoscale regions with the lowest $T_c$. This local switching will produce complex strain fields that could drive complex cooperative effects[42], which could now be directly visualized. Furthermore, as the 0.25 eV of X-ray bandwidth used here corresponds to a transform limited pulse duration of less than 20 fs, decoupling of the bandgap changes from structural dynamics could be resolved during the sub-100 fs initial evolution of the system[2]. Our approach can easily be extended to characterize filament growth in electric fields, which will enable spectroscopy of field-induced states driven by quasi-d.c. fields or THz pulses[9,43]; thus, heterogeneous transient states in quantum materials can now be explored at the nanoscale.

## Online content

Any methods, additional references, Nature Portfolio reporting summaries, source data, extended data, supplementary information, acknowledgements, peer review information; details of author contributions and competing interests; and statements of data and code availability are available at https://doi.org/10.1038/s41567-022-01848-w.

## Methods

### Sample preparation and measurements

Samples consisted of 75-nm-thick layers of $VO_2$ prepared on silicon nitride membranes by pulsed laser deposition and subsequent annealing, as widely used to prepare high-quality $VO_2$ films[8,9,32,33,44]. A [Cr(5 nm)/Au(55 nm)]20 multilayer (~1.1 µm thickness) was deposited on the opposite side, and a focused ion beam was used to mill the mask structure. A 2-µm-diameter aperture was milled to define the field of view, along with five 50- to 90-nm-diameter reference apertures 5 µm away from the central aperture. Crystals grow with the c axis preferentially in plane.

Pump–probe experiments were performed at the FTH end station of the soft x-ray spectroscopy and scattering beamline at the PAL-XFEL operating at 60 Hz repetition rate. The X-ray polarization was perpendicular to the rutile c axis. Time-resolved images were taken by alternating between positive and negative time delays to ensure that the initial domain structure did not change. Each image averages 9,000 X-ray free electron laser (XFEL) shots. The XFEL beam was focused to a 50 µm × 50 µm spot size at the sample position using a Kirkpatrick–Baez mirror pair. To prevent the sample from being damaged, the XFEL was attenuated to have ~$1.2 \times 10^8$ photons per pulse. Scattering patterns were captured by an in-vacuum charge-coupled device detector (PI-MTE 2048b, Teledyne Princeton Instruments) cooled to −40 °C and read out at 100 kHz with 2 × 2 binning. The charged-coupled device was placed 300 mm downstream from the sample, with a metallic filter (100 nm parylene/100 nm aluminium from Luxel) and 1000-µm-thick epoxy beamstop placed directly before the sensor. Optical pump-only backgrounds were subtracted from each image before either iterative reconstruction or holographic inversion. Laser pulses with a central wavelength of 800 nm were focused to a spot size of 200 µm full-width at half-maximum (FWHM) at the sample with a small (~1°) crossing angle relative to the normal incidence X-rays. The overall temporal resolution was around 150 fs, limited by the relative timing jitter of the optical and X-ray beams. Both beams impinged on the sample $VO_2$ side first to reduce the potential impact of plasmonic effects in the Cr/Au multilayer introducing additional inhomogeneity[3].

### FTH and CDI

In FTH, a beam block can be added to the scattering field to block the intense low-momentum scatter, allowing single exposures to capture the high-momentum scattering needed to observe the nanoscale domains. However, the presence of beam block in the central part of the Fourier plane removes the d.c. component and acts as a high-pass filter. As a result, the absolute values of the complex transmission of the sample are lost. In CDI, multiple exposures are combined to improve dynamic range, enabling better sampling of both low- and high-angular scattering. Although, in principle, the beam block can be removed to capture the d.c. component, it is not necessary for CDI to reconstruct the d.c. level if the X-ray beam has sufficient coherence, which is the case at the FEL. Instead, multiple exposures with the beam block in place were sufficient to produce good CDI results via iterative phase retrieval algorithms[33].

### Principal component analysis and data fitting

The dynamics of the real space images were analysed using principal component (PC) analysis. First, all negative time delays (probe before pump images) from each time trace were grouped and decomposed into PCs; as these images should be constant over time, the amplitude of the second PC was used to define a noise threshold for the pump–probe measurements. Then each of the two time traces (0–1 ps and 0–20 ps) was decomposed into PCs, and any PC with an amplitude below the noise threshold was discarded. Only two PCs were found to be significant for each time trace: one with a spatial pattern resembling the negative time delay structure with a weak structure in time and one approximately constant at negative time delays and exhibiting a sharp drop at temporal overlap. We identified the first as the static background, with dynamics resulting only from fluctuations in the overall XFEL brightness, and the second as the dynamics of the phase transition. We replaced the first component with its time-averaged mean value and kept the second. We note that including the other, weaker PCs barely affects the time traces; therefore, we ascribed all time dynamics to the second principal component (Extended Data Fig. 5). The remaining time dynamics can be described as a purely real function. Including images taken at longer time delays (100 ps, 250 ps or 500 ps) breaks this simple description, and additional PCs are required to describe the dynamics, corresponding to the onset of spatial dynamics (Extended Data Fig. 4).

While the short and long time traces cannot be directly combined because of a change in the initial domain structure, which we attribute to a random fluctuation in the XFEL intensity introducing irreversible changes, examination of PC corresponding to the dynamics shows they are the same in both cases (Extended Data Fig. 6). As such, we fit the time dependent signal, $S$, of both traces simultaneously with a double exponential decay of the form $S(t) = H(t)\left(A \exp\left(-\frac{t}{\tau_1}\right) + B \exp\left(-\frac{t}{\tau_2}\right) + C\right) + 1$, where $H(t)$ is the Heaviside step function, with identical time constants ($\tau_1, \tau_2$) but freely varying amplitudes ($A, B, C$) and convolved with the time resolution (150 fs FWHM). Error bars on the fit represent the standard deviation of the fit assuming a standard deviation of 0.03% on the data, obtained from examining the pre-time-zero data. Uncertainty in the time resolution, pump fluctuations and penetration depth mismatch are not explicitly considered, but increase uncertainty at the femtosecond timescale.

### Reconstruction of the transient phase

Full amplitude and phase images of transient phase were recovered using partially coherent iterative phase reconstruction algorithms as described in ref. [33]. Long and short exposures were combined to provide high dynamic range images at 31 photon energies across the oxygen K edge, which were each used to reconstruct real space images independently. The object constraint was determined from the known mask geometry and the FTH reconstructions; a data constraint mask was used to allow the reconstruction to freely vary in regions where the beamblock masked the detector or where additional background light or detector damage was present, with the additional constraint that the blocked low-$q$ response obeyed circular symmetry. The properties of the transient state were determined by comparing the average spectral response of all sample regions that began in the R phase and those that were switched from the M phase. These regions were determined by combining FTH images at positive and negative time delays at three different photon energies; only regions where all three photon energies agreed were preserved for further analysis. These regions also agree clearly with a PC analysis of the spectrogram, although the degree of sample inhomogeneity prevented a conventional clustering analysis.

### XAS simulations

XAS simulations were performed using density functional theory (DFT) and the supercell core–hole[45] method as implemented in Vienna Ab initio Simulation Package[46] with projector-augmented wave[47] pseudopotentials. For all calculations, supercells containing 16 $VO_2$ formula units (48 atoms) were employed, and 800 bands were included in the core–hole calculations. The plane-wave cut-off was 600 eV, and a 4 × 4 × 6 $k$-point mesh was used. XAS absorption was approximated by the imaginary part of the frequency-dependent dielectric function, which was convoluted with a Gaussian function with a FWHM of 0.3 eV and a Lorentzian function that linearly increases in width from 0.0 to 0.7 eV between 520 eV and 540 eV to simulate the instrument and lifetime broadening. The simulated spectra of the undistorted $VO_2$ were manually aligned with experiment, and the same energy shift was applied to the distorted structures. We compared three DFT methods,

LDA, PBE and PBE+$U$ with a rotationally invariant Hubbard-$U$ correction of 3.1 eV for the vanadium $d$ bands. The three methods did not show any noteable differences for the oxygen K edge (Extended Data Fig. 8). Reported simulations are for X-ray polarization perpendicular to the rutile $c$ axis and were based on LDA and are in good agreement with previously published measurements for VO$_2$ (refs. [33,48,49]), although the amplitude of σ* state at 532 eV can be seen to vary between different measurements and calculation. The undistorted rutile structure was calculated with lattice parameters $a_R = b_R = 4.5546$ Å and $c_R = 2.8514$ Å, reported in ref. [50]. The orthorhombic phase is calculated by projecting the monoclinic axes onto their rutile counterparts while removing the dimerization and tilt distortions, so that the vanadium ions are evenly spaced along each axis, that is, the vanadium ions are located at (0, 0, 0) and (0.5, 0.5, 0.5) in fractional coordinates of the unit cell, as in the rutile phase. This orthorhombic structure has lattice parameters $a_O = b_M = 4.5378$ Å, $b_O = c_M \sin \beta_M = 4.5322$ Å, and $c_O = a_M/2 = 2.8759$ Å, where $a_M, b_M, c_M$ and $\beta_M$ are the monoclinic unit cell lengths and angles reported in ref. [51]. At the measurement time of 20 ps, the out-of-plane strain has relaxed, so we finally set the orthorhombic $b_O$ axis, which lies out of plane, to the rutile value to give $a_O = 4.5378$ Å, $b_O = 4.5546$ Å, and $c_O = 2.8759$ Å.

## Data availability

All data shown in this study are included in this published article (and its Supplementary Information files). Raw holograms are available from the corresponding authors on reasonable request. Source data are provided with this paper.

## Acknowledgements

We acknowledge the support of J. Turner and A. Reid with an early iteration of this experiment. This work was funded through the European Research Council (ERC) under the European Union's Horizon 2020 Research and Innovation Programme (grant agreement no. 758461) and PGC2018-097027-B-I00 project funded by MCIN/AEI/10.13039/501100011033/FEDER 'A way to make Europe' and CEX2019-000910-S (MCIN/AEI/10.13039/501100011033), Fundació Cellex, Fundació Mir-Puig, and Generalitat de Catalunya (AGAUR grant no. 2017 SGR 1341, CERCA programme) and by the Independent Research Fund Denmark under the Sapere Aude programme (grant no. 9064-00057B) and VILLUM FONDEN under the Young Investigator Program (grant no. 15375). S.Kim., S.C and H.K. acknowledge support from the National Research Foundation of Korea (NRF-2021R1A3B1077076). S. Kwon acknowledges support from the National Research Foundation of Korea (NRF-2020R1A2C1007416). Part of this work was carried out with the support of Diamond Light Source, instrument I06 (proposal MM22048). K.H. and R.F.H. were supported by the US National Science Foundation (EECS-1509740). N. Artrith thanks the Dutch National e-Infrastructure and the SURF Cooperative for computational resources that were used for the DFT XAS simulations. A.S.J. acknowledges support of a fellowship from 'la Caixa' Foundation (ID 100010434), fellowship code LCF/BQ/PR21/11840013, and support from the Marie Skłodowska-Curie grant agreement no. 754510 (PROBIST) and the Agencia Estatal de Investigacion (the R&D project CEX2019-000910-S, funded by MCIN/AEI/10.13039/501100011033, Plan National FIDEUA PID2019-106901GB-I00, FPI).

## Author contributions

A.S.J., D.P.-S., K.M.S., S.Kim., S.C. and S.H.P. performed experiments at PAL-XFEL and were remotely assisted by K.V., P.E.M. and S.E.W. The beamline and scattering chamber were designed and built by A.K., S.Kwon. and S.H.P. The laser was run by H.C. and D.J. Data analysis and software development was led by A.S.J. with help from D.P.-S. Samples were prepared and characterized by K.H. and R.F.H. and processed for holography by C.M.G. Further characterization of the samples was performed at the ALBA Synchrotron with collaboration by A.S.J. and D.P.-S. with P.G. and M.V. and remote support from K.V. and P.E.M. Domain stability measurements were performed at the Diamond Light Source by A.S.J. and D.P.-S. with support from D.B., F.M., A.F. and S.D. N. Artrith, N. Agarwal and F.d.G. performed calculations of the XAS spectra. The project was conceived by S.E.W. and planned and designed by A.S.J., S.U., B.P., S.E., H.K. and S.E.W. A.S.J., D.P.-S. and S.E.W. wrote the manuscript with input from all authors.

## Competing interests

The authors declare no competing interests.

## Additional information

**Extended data** is available for this paper at https://doi.org/10.1038/s41567-022-01848-w.

**Supplementary information** The online version contains supplementary material available at https://doi.org/10.1038/s41567-022-01848-w.

**Correspondence and requests for materials** should be addressed to Allan S. Johnson or Simon E. Wall.

**Peer review information** *Nature Physics* thanks the anonymous reviewers for their contribution to the peer review of this work.

**Reprints and permissions information** is available at www.nature.com/reprints.

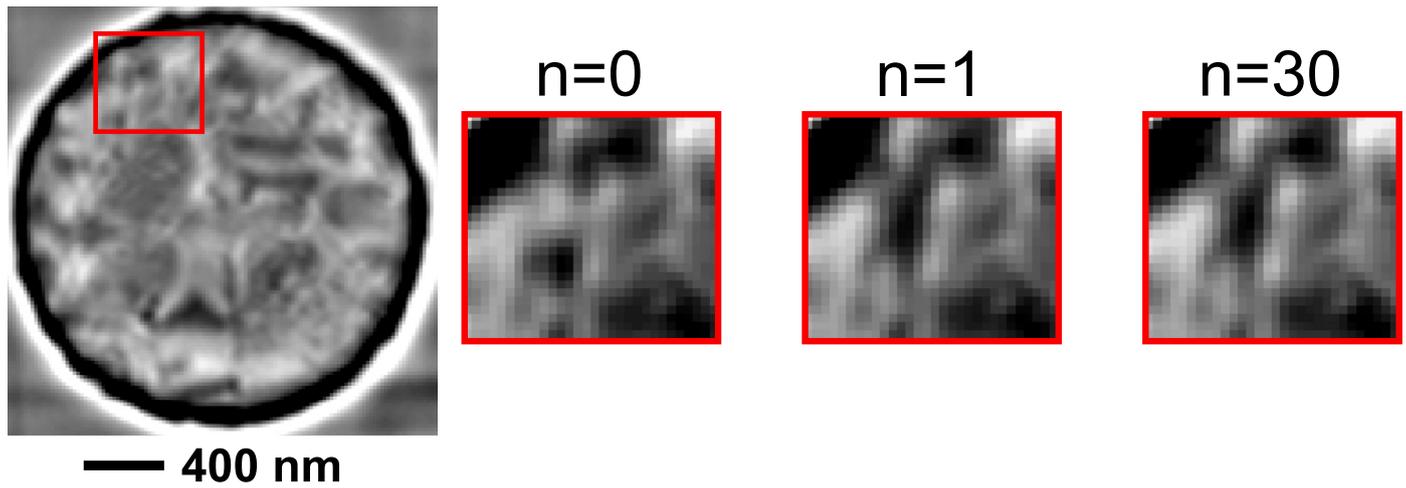

**Extended Data Fig. 1 | Laser induced domain changes.** Images taken at the Diamond light source of a similar $VO_2$ sample at 529.5 eV photon energy after repeated exposure to 33 mJ/cm$^2$ excitation. After one exposure (n = 1) the system relaxes to a new domain structure, but repeated exposures (n = 30) to the same fluence do not further change the laser induced domain structure.

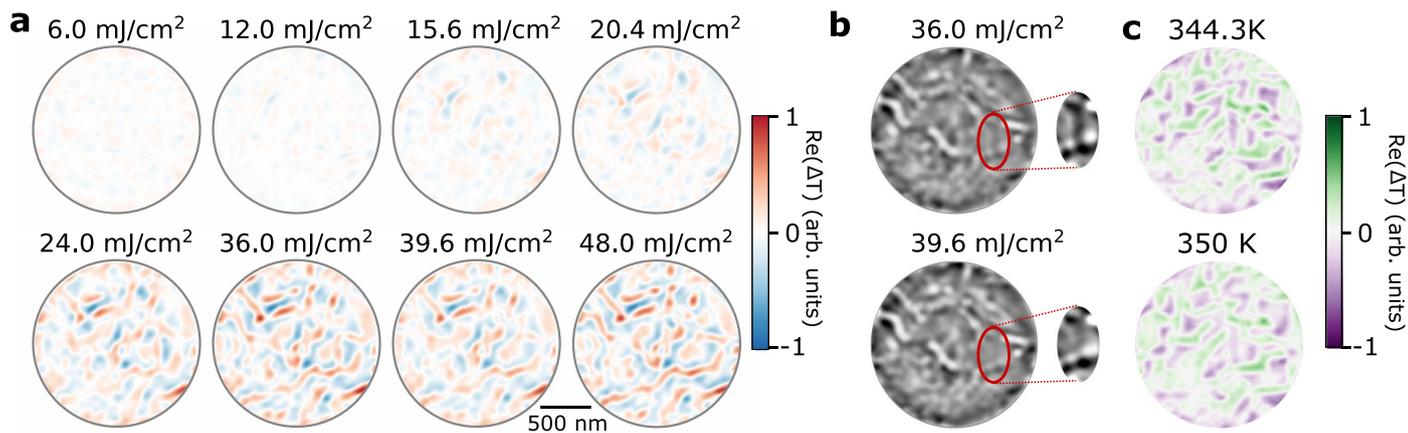

**Extended Data Fig. 2 | Fluence dependence of the transient dynamics.**
**a** Differential signal at 20 ps time delay and 529.5 eV photon energy as a function of fluence. As the fluence increases more domains are switched, until changes rise rapidly around 20 mJ/cm² and saturate. This saturation is associated to switching all of the domains and the full depth of the bulk. Changes above 24 mJ/cm² are associated with changes in the initial state rather than dynamics, as shown in panel b. **b** Pre-time-zero images at 36 mJ/cm² and 39.6 mJ/cm², as used in part a. When the fluence is increased the initial domain pattern undergoes an irreversible change. For example, the highlighted region shows the erasure of a vertical stripe domain. This is attributed to X-ray shots with higher than average flux[52] **c** Thermal domain structure measured at the ALBA synchrotron. The domain pattern is similar but exhibits some notable differences, attributable to the rapid cooling cycles in the time-resolved experiment.

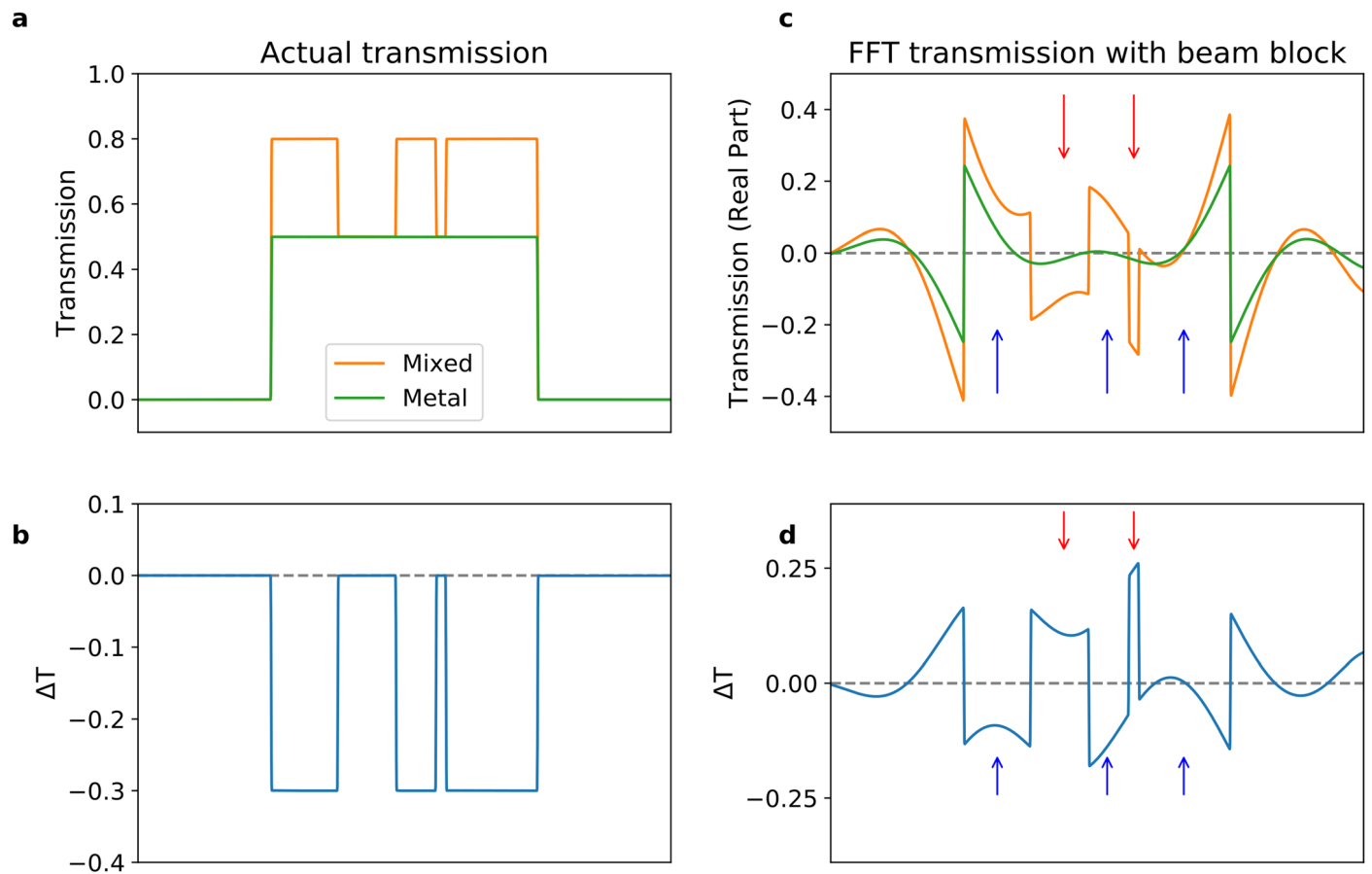

**Extended Data Fig. 3 | Demonstration of the effects of the beam block on FTH images. a** Hypothetical transmission of a sample in the case of a homogenous metal (green) and a mixed-phase (orange) of metal and insulating regions (insulating regions have a higher transmission). **b** Change in transmission when going from the mixed-phase to the metallic phase; only the insulating phase shows changes. **c** FTH image corresponding to the transmission function in panel **a**. **d** Change in transmission for the FTH image, in this case both insulating (indicated by the blue arrows) and metallic (red arrows) can show an apparent change.

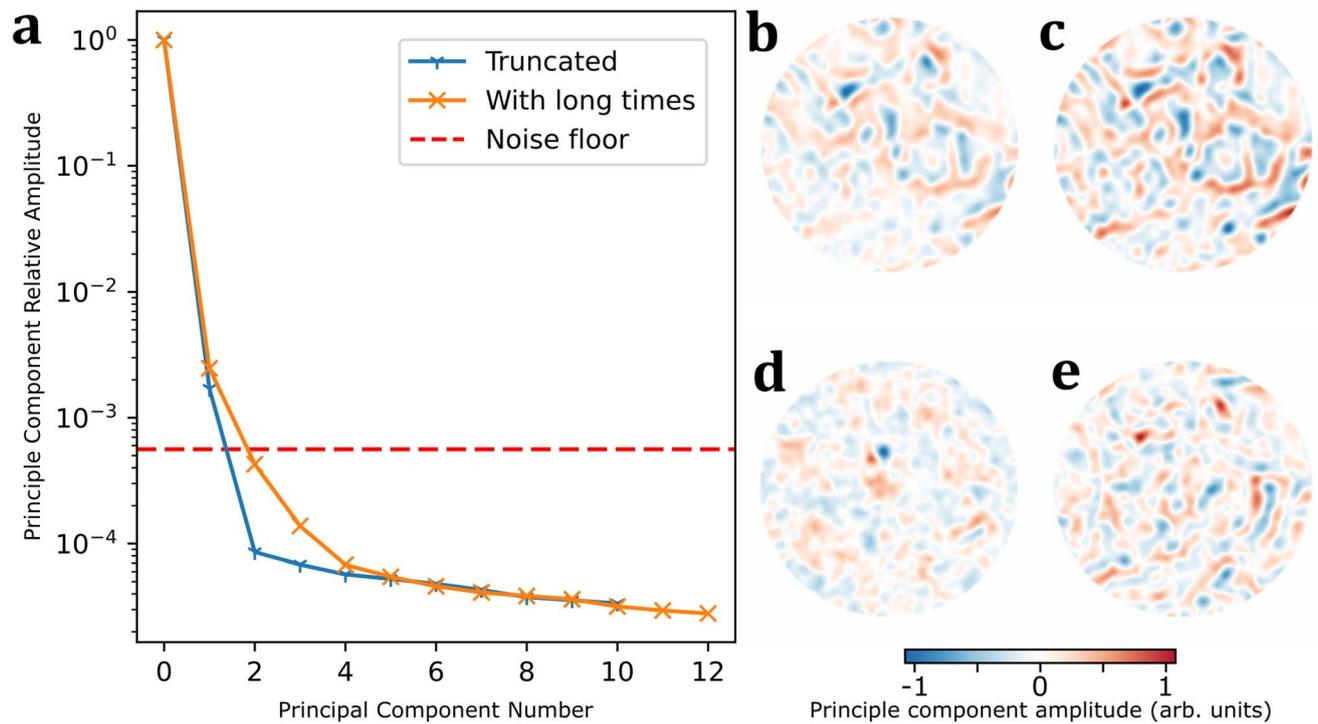

**Extended Data Fig. 4 | Amplitude of the principal components. a** Relative amplitude of the principal components considering only data in the first picosecond (truncated) and including data at 250 and 500 ps (with long times). Also shown is the noise floor. **b,d** PC1 (top) and PC2 (bottom) using truncated time data. **c,e** PC1 and PC2 including data out to 500 ps.

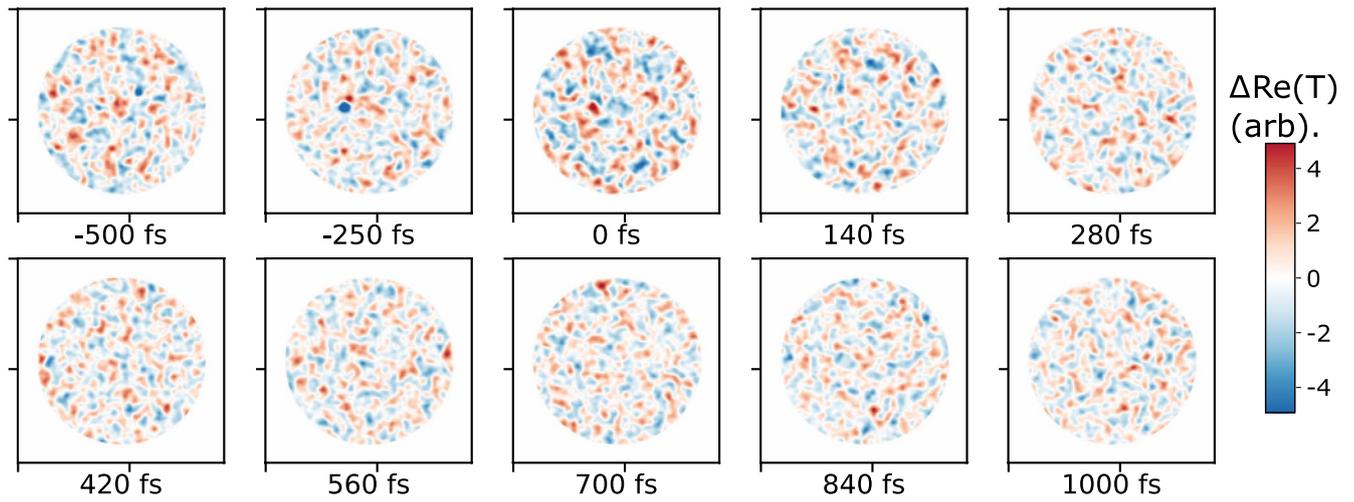

**Extended Data Fig. 5 | Residual signal after subtracting the two main principal components.** No systematic changes are observed, and all remaining dynamics are compatible with noise.

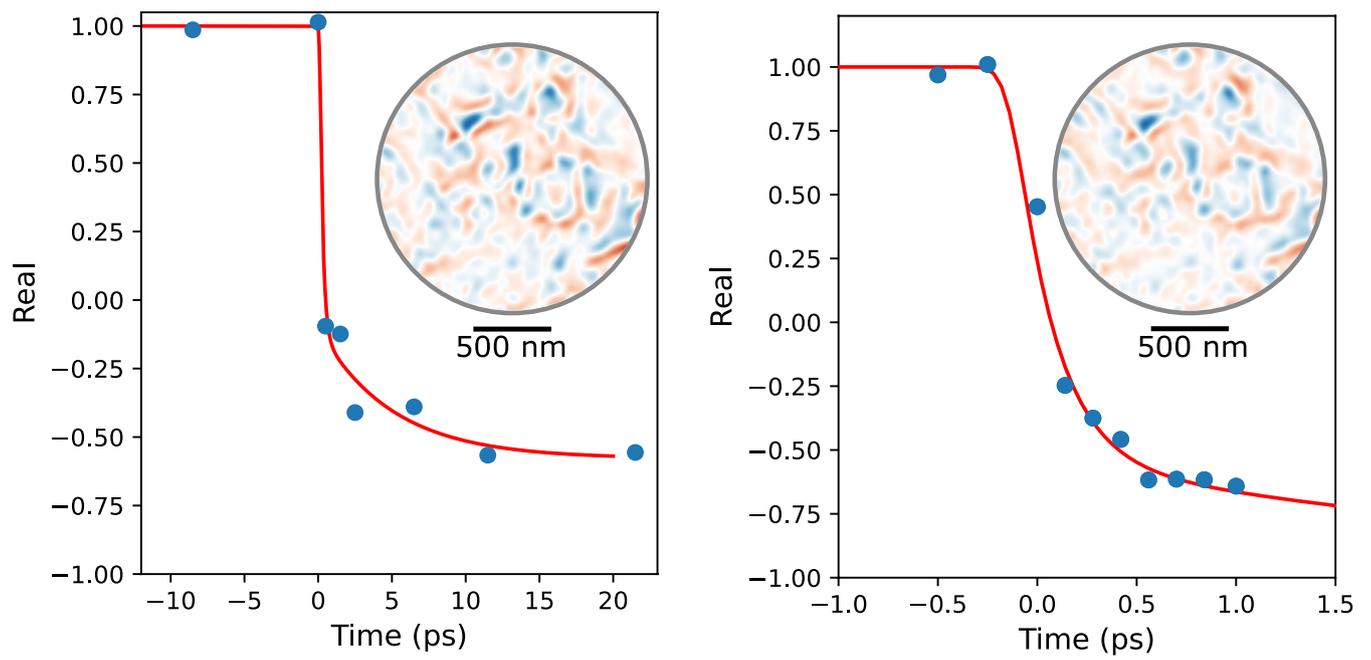

**Extended Data Fig. 6 | Comparison of short and long time data.** Left, the long-time scan data used for fitting the slow 4.5 ps time constant. Right, reproduction of the short time data shown in Fig. 2 of the main manuscript. Insert shows the spatial dependence of the principal component.

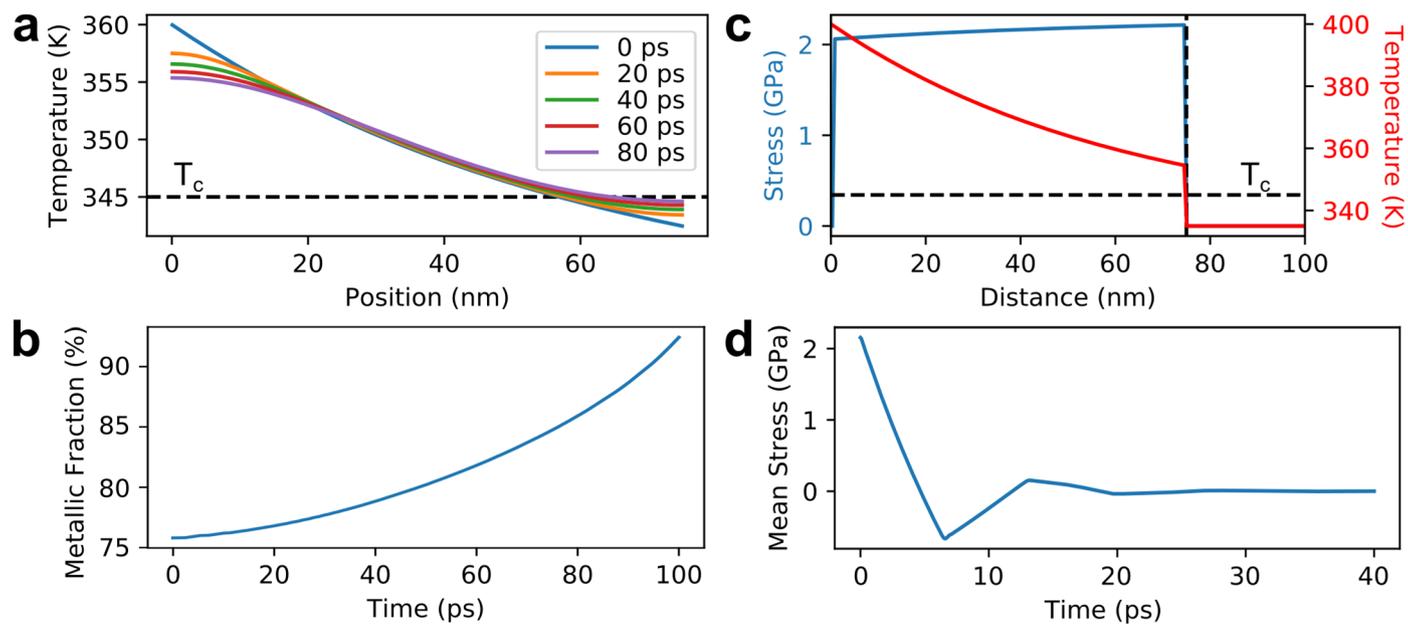

**Extended Data Fig. 7 | Simulated thermal and acoustic propagation in VO$_2$ thin films.** a Temperature profile in the sample as a function of time for below threshold excitation. b Corresponding metallic fraction vs time. c Initial stress field from a homogeneously switched film (above threshold excitation). d Resulting stress relaxation dynamics in the out-of-plane direction. Calculations made use of tabulated data in ref. [53].

 

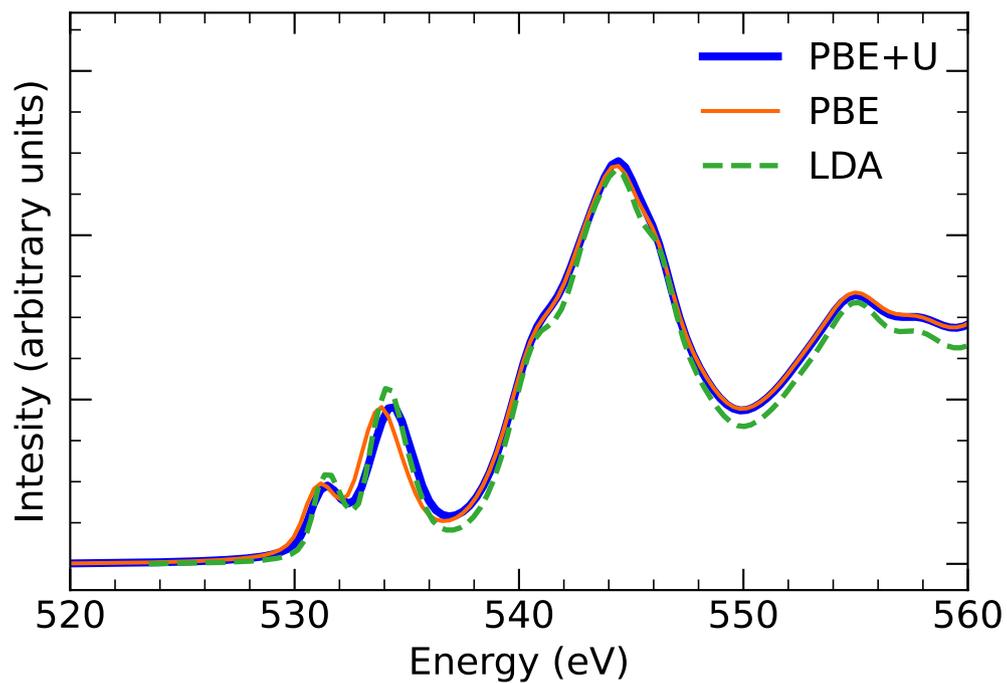

**Extended Data Fig. 8 | DFT calculations of the oxygen K-edge in $VO_2$.** XAS lines obtained from calculations with LDA, PBE, and PBE + U are compared. The spectra were visually aligned. The simulated lines were convoluted with a Gaussian of 0.3 eV (FWHM) and a Lorentzian that linearly increases in width from 0.0 to 5.0 eV between 520 eV and 560 eV to simulate the instrument and lifetime broadening and facilitate comparison to literature spectra.